\documentclass[aps,prd,onecolumn,groupedaddress,showpacs,nofootinbib,amssymb]{revtex4}
\usepackage[dvips]{graphicx}
\usepackage{amssymb}
\usepackage{amsmath}
\usepackage{graphicx,,color}
\usepackage{amsfonts}
\usepackage{bm}
\usepackage{cancel}
\usepackage{comment}

\newcommand\be{\begin{equation}}
\newcommand\ee{\end{equation}}
\newcommand\nn{\nonumber \\}
\newcommand\e{\mathrm{e}}

\allowdisplaybreaks[4]

\begin{document}

\title{Constant-roll Inflation in $F(R)$ Gravity}
\author{S.~Nojiri,$^{1,2}$\,\thanks{nojiri@gravity.phys.nagoya-u.ac.jp}
S.~D.~Odintsov,$^{3,4}$\,\thanks{odintsov@ieec.uab.es}
V.~K.~Oikonomou,$^{5,6}$\,\thanks{v.k.oikonomou1979@gmail.com}}
\affiliation{$^{1)}$ Department of Physics, Nagoya University,
Nagoya 464-8602, Japan \\
$^{2)}$ Kobayashi-Maskawa Institute for the Origin of Particles and
the Universe, Nagoya University, Nagoya 464-8602, Japan \\
$^{3)}$ ICREA, Passeig Luis Companys, 23, 08010 Barcelona, Spain\\
$^{4)}$ Institute of Space Sciences (IEEC-CSIC) C. Can Magrans s/n,
08193 Barcelona, Spain\\
$^{5)}$ Laboratory for Theoretical Cosmology, Tomsk State University
of Control Systems
and Radioelectronics, 634050 Tomsk, Russia (TUSUR)\\
$^{6)}$ Tomsk State Pedagogical University, 634061 Tomsk, Russia\\
}

\tolerance=5000

\begin{abstract}
We propose the study of constant-roll inflation in $F(R)$ gravity.
We use two different approaches, one that relates an $F(R)$ gravity
to well known scalar models of constant-roll and a second that
examines directly the constant-roll condition in $F(R)$ gravity.
With regards to the first approach, by using well known techniques,
we find the $F(R)$ gravity which realizes a given constant-roll
evolution in the scalar-tensor theory. We also perform a conformal
transformation in the resulting $F(R)$ gravity and we find the
Einstein frame counterpart theory. As we demonstrate, the resulting
scalar potential is different in comparison to the original scalar
constant-roll case, and the same applies for the corresponding
observational indices. Moreover, we discuss how cosmological
evolutions that can realize constant-roll to constant-roll eras
transitions in the scalar-tensor description, can be realized by
vacuum $F(R)$ gravity. With regards to the second approach, we
examine directly the effects of the constant-roll condition on the
inflationary dynamics of vacuum $F(R)$ gravity. We present in detail
the formalism of constant-roll $F(R)$ gravity inflationary dynamics
and we discuss how the inflationary indices become in this case. We
use two well known $F(R)$ gravities in order to illustrate our
findings, the $R^2$ model and a power-law $F(R)$ gravity in vacuum.
As we demonstrate, in both cases the parameter space is enlarged in
comparison to the slow-roll counterparts of the models, and in
effect, the models can also be compatible with the observational
data. Finally, we briefly address the graceful exit issue.
\end{abstract}

\pacs{04.50.Kd, 95.36.+x, 98.80.-k, 98.80.Cq,11.25.-w}

\maketitle

\section{Introduction \label{SecI}}

The inflationary paradigm is one of the most widely accepted
scenarios that describes the early Universe evolution.
Traditionally, the description of the inflationary era is given in
terms of a slow-rolling single scalar field, and many reviews
already exist in the literature that describe the single scalar
inflation \cite{Linde:2007fr,Gorbunov:2011zzc,Lyth:1998xn}. The
observational data coming from Planck 2015 \cite{Ade:2015lrj}
restricted quite significantly the single scalar field inflationary
models, and many models were rendered non-viable
\cite{Martin:2016ckm}. However some single scalar models remained
viable after the Planck constraints were imposed on them, and
actually these models have quite appealing properties, such as the
Starobinsky model \cite{Starobinsky:1980te,Barrow:1988xh}, the Higgs
model \cite{Bezrukov:2007ep}, and a wide class of models called
$\alpha$-attractors
\cite{Kallosh:2013hoa,Ferrara:2013rsa,Kallosh:2013yoa}, see also
\cite{Odintsov:2016vzz,Odintsov:2016jwr} for the $F(R)$ gravity
realization of $\alpha$-attractors.

Despite of the appealing properties of the single scalar field
models of inflation, these models have a potential drawback related
to non-Gaussianities \cite{Chen:2010xka}. Particularly, up to date
the modes of the spectrum of the primordial curvature perturbations
are assumed to be uncorrelated, so the spectrum is assumed to obey a
Gaussian distribution. However, if non-Gaussianities are observed in
the future, then the single scalar field models will be put into
question since these do not predict any non-Gaussianities. One
conceptually appealing way to introduce non-Gaussianities in single
scalar field models is to directly modify the slow-roll condition.
This approach was used in
Refs.~\cite{Inoue:2001zt,Tsamis:2003px,Kinney:2005vj,Tzirakis:2007bf,
Namjoo:2012aa,Martin:2012pe,Motohashi:2014ppa,Cai:2016ngx,
Motohashi:2017aob,Hirano:2016gmv,Anguelova:2015dgt,Cook:2015hma,
Kumar:2015mfa,Odintsov:2017yud,Odintsov:2017qpp}, see also
\cite{Lin:2015fqa,Gao:2017uja} for an alternative viewpoint. The
models used in
Refs.~\cite{Inoue:2001zt,Tsamis:2003px,Kinney:2005vj,Tzirakis:2007bf,
Namjoo:2012aa,Martin:2012pe,Motohashi:2014ppa,Cai:2016ngx,
Motohashi:2017aob,Hirano:2016gmv,Anguelova:2015dgt,Cook:2015hma,
Kumar:2015mfa,Odintsov:2017yud,Odintsov:2017qpp} are known as
constant-roll models, and in all the cases the slow-roll era is
modified. In effect, a non-zero amount of non-Gaussianities may
appear in the primordial power spectrum
\cite{Namjoo:2012aa,Martin:2012pe}.

In this paper we aim to find the $F(R)$ gravity description (for
reviews on $F(R)$ gravity see
\cite{Nojiri:2006ri,Capozziello:2011et,Capozziello:2010zz,
Capozziello:2009nq,Nojiri:2010wj,Clifton:2011jh}) of the models of
constant-roll evolution. Our approach is two-fold, since firstly we
shall realize the Hubble rate of some constant-roll models by using
a vacuum $F(R)$ gravity. After finding the $F(R)$ gravity, we shall
perform a conformal transformation in order to obtain the Einstein
frame theory, and as we show, the obtained Einstein frame theory is
different in comparison to the constant-roll models. However, it is
possible to obtain concordance with the observations even for the
obtained Einstein frame theories, and we calculate the spectral
index and the scalar-to-tensor ratio. Also we briefly investigate
which $F(R)$ gravity can realize a model which is known to generate
transitions between constant-roll eras in the Einstein frame. This
difference between the Einstein frame and the $F(R)$ gravity was
also noticed in the literature, see for example
\cite{Domenech:2016yxd,Bahamonde:2017kbs,Bahamonde:2016wmz,Brooker:2016oqa}.
Our second approach to the constant-roll problem in the context of
$F(R)$ gravity is more direct in comparison to our first approach,
since we shall find the implications of the constant-roll condition
directly in the $F(R)$ gravity theory, without invoking the
scalar-tensor theories and their Hubble rate. This approach is more
direct for the reason that the implications of the constant-roll
condition can be seen in detail in the qualitative features of the
$F(R)$ gravity. Particularly, we shall investigate how the
constant-roll condition affects the inflationary indices used for
the study of inflation, and we also calculate the spectral index of
primordial curvature perturbations and the scalar-to-tensor ratio.
In order to demonstrate the implications of the constant-roll
condition in $F(R)$ gravity, we shall use two well-known models, the
$R^2$ model and a power-law $F(R)$ gravity model. In both cases we
shall make two crucial assumptions, firstly that the first slow-roll
index $\epsilon_1=-\frac{\dot{H}}{H^2}$ is very small during the
inflationary era, a condition that was also assumed to hold true in
Ref.~\cite{Martin:2012pe}. Secondly, we shall assume that although
$\epsilon_1\ll 1$, the constant-roll condition holds true. As we
shall demonstrate, in the case of the constant-roll $R^2$ model, the
constant-roll condition affects the resulting qualitative features
of the model, making it compatible with the 2015-Planck
\cite{Ade:2015lrj} and BICEP2/Keck-Array data \cite{Array:2015xqh},
for a wider range of the parameter space in comparison to the
ordinary $R^2$ model. We also discuss in brief some restrictions and
drawbacks of the constant-roll approach in the case of the $R^2$
model, which possibly occur due to the lack of analyticity in our
approach. Finally, we briefly address the graceful exit issue and we
investigate which restrictions it imposes on the parameter space. We
perform the same analysis for the power-law $F(R)$ gravity model,
and as we demonstrate, the constant-roll power-law $F(R)$ gravity
model can be compatible with the current observational data, in
contrast to the slow-roll model which is not compatible with the
observations. We need to note that in both the $F(R)$ gravity models
we shall study, if the constant-roll condition is canceled, the
results of the constant-roll case coincide with the slow-roll case,
a behavior possibly expected since we assumed that $\epsilon_1\ll
1$.

This paper is organized as follows: In section \ref{SecII}, we
present the formalism with which we will be able to find the $F(R)$
gravity description of certain scalar-tensor constant-roll models.
By using the formalism, we shall realize a particularly interesting
scenario of constant-roll inflation. Also we shall find the
corresponding Einstein frame picture and we shall show that
concordance with the observations may be achieved. In addition we
present the $F(R)$ gravity which realized a cosmic evolution that in
the scalar field frame is known to produce transitions between
constant-roll eras. In section \ref{SecIII}, we adopt a more direct
approach, and we investigate the implications of the constant-roll
condition in a vacuum $F(R)$ gravity. We present the formalism of
the constant-roll $F(R)$ gravity and we find explicit expressions
for the inflationary indices, by also comparing the results with the
slow-roll case. In order to illustrate our findings, we present the
constant-roll $R^2$ model, and also a well known power-law $F(R)$
gravity model, and we discuss the implications and the shortcomings
of the constant-roll approach, by also comparing the results with
the slow-roll case. Finally the conclusions along with a discussion
follow in the end of the paper.

Before we start our presentation, let us here briefly discuss the
geometric conventions we shall assume to hold true for the rest of
this paper. We shall assume that the background metric is a flat
Friedmann-Robertson-Walker (FRW) metric with line element,
\begin{equation}
\label{metricfrw} ds^2 = - dt^2 + a(t)^2 \sum_{i=1,2,3}
\left(dx^i\right)^2\, ,
\end{equation}
where $a(t)$ denotes as usual the scale factor. Also, we assume that
the metric connection is the Levi-Civita connection, which is an
affine connection which is, metric compatible, torsion-less and
symmetric.

\section{$F(R)$ constant-roll Inflation and Einstein Frame \label{SecII}}

The constant-roll inflation scenario was introduced as an
alternative to the slow-roll scenario, with the first having the
appealing feature that non-Gaussianities are generated. We shall
consider one model which was introduced in
Refs.~\cite{Motohashi:2014ppa, Motohashi:2017aob}, and it was
studied in the context of scalar field theory. In this paper we
shall be interested in realizing the resulting cosmic evolution of
the scalar model appearing in
Refs.~\cite{Motohashi:2014ppa,Motohashi:2017aob}, in the context of
$F(R)$ gravity, and we explicitly construct a model which realizes
the constant-roll inflation model. As it is well-known, by using a
conformal transformation, the action of the $F(R)$ gravity can be
rewritten as a scalar-tensor theory. As we demonstrate, the $F(R)$
gravity which realizes the model of constant-roll when it is
conformally transformed in the Einstein frame, the resulting
scalar-tensor theory is different from that used in
\cite{Motohashi:2014ppa, Motohashi:2017aob}. The spectral index
$n_s$ of primordial curvature perturbations and the scalar-to-tensor
ratio $r$ can be calculated by using the resulting Einstein frame
theory. According to the cosmological observations, these quantities
can be obtained from the correlation function of the density with
respect to the angle. The angle is unaffected by the conformal
transformation, and the amplitude is changed by a factor which does
not depend on the spacial coordinates if we consider isotropic
background. This indicates that the quantities could be obtained by
using the potential in the scalar-tensor theory obtained from the
$F(R)$ gravity theory. These quantities should be different from
those obtained in \cite{Motohashi:2014ppa,Motohashi:2017aob} since
the potentials in the scalar-tensor theory obtained from the $F(R)$
gravity are different from the ones corresponding to the
scalar-tensor theory \cite{Motohashi:2014ppa, Motohashi:2017aob}.

We begin by considering the general solution for the constant-roll
condition in the context of scalar-tensor theory,
\begin{equation}
\label{cr1} \ddot \phi = \beta H \dot \phi \, .
\end{equation}
By using the FRW equations,
\begin{equation}
\label{cr2} \frac{3}{\kappa^2} H^2 = \frac{1}{2}{\dot\phi}^2 +
V(\phi)\, , \quad
 - \frac{1}{\kappa^2} \left( 3 H^2 + 2 \dot H \right)
= \frac{1}{2}{\dot\phi}^2 - V(\phi)\, ,
\end{equation}
we obtain,
\begin{equation}
\label{cr3} \frac{2}{\kappa^2} \dot H= {\dot \phi}^2 \, .
\end{equation}
Then we acquire,
\begin{equation}
\label{cr4} \frac{2}{\kappa^2} \ddot H= 2 \dot\phi \ddot \phi \, .
\end{equation}
By eliminating $\ddot \phi$ and $\dot \phi$ by using
Eqs.~(\ref{cr1}), (\ref{cr3}), (\ref{cr4}) we obtain,
\begin{equation}
\label{cr5} 0 = \ddot H - 2\beta H \dot H \, ,
\end{equation}
which can be integrated as follows,
\begin{equation}
\label{cr6} \dot H - \beta H^2 = C \quad \left(\mbox{constant}
\right)\, .
\end{equation}
We may furthermore rewrite Eq.~(\ref{cr6}) in the following form,
\begin{equation}
\label{cr7} \frac{d^2 a^{-\beta}}{dt^2} = - \beta C a^{-\beta} \, ,
\end{equation}
When $\omega^2 \equiv \beta C > 0$, the solution of (\ref{cr7}) is
given by,
\begin{equation}
\label{cr8} a^{-\beta} = A \cos \omega t + B \sin \omega t \, .
\end{equation}
On the other hand, if $- \lambda^2 \equiv \beta C <0$, we find
\begin{equation}
\label{cr9} a^{-\beta} = C \cosh \lambda t + D \sinh \lambda t \, ,
\end{equation}
where $A$, $B$, $C$, and $D$ are arbitrary constants.

In the following, we consider the following cosmological evolution
which was introduced in
Refs.~\cite{Motohashi:2014ppa,Motohashi:2017aob},
\begin{equation}
\label{cr10} H=-M \tanh \left( \beta M t \right) \quad \left( a
\propto \cosh^{-\frac{1}{\beta}} \left( \beta M t \right) \right)\,
.
\end{equation}
In order to construct the $F(R)$ gravity model which reproduces
Eq.~(\ref{cr10}),  we rewrite the action of the $F(R)$ gravity as
follows, \cite{Nojiri:2006gh},
\begin{equation}
\label{cr11} S= \frac{1}{2\kappa^2} \int d^4 x \sqrt{-g} \left(
P(\phi) R + Q(\phi) \right)\, .
\end{equation}
By varying the action (\ref{cr11}) with respect to the auxiliary
scalar field $\phi $, we obtain the equation,
\begin{equation}
\label{auxiliaryeqns} P'(\phi )R+Q'(\phi)=0\, ,
\end{equation}
which can be solved with respect $\phi$ as a function of the scalar
curvature $R$ as $\phi (R)$. Then by substituting in the initial
action, we obtain the $F(R)$ gravity whose Lagrangian density is
given by
\begin{equation}
\label{r1} F(\phi( R))= P (\phi (R))R+Q (\phi (R))\, .
\end{equation}
Since we can redefine the scalar field freely, we may identify the
scalar field $\phi$ with the time coordinate $t$, $\phi=t$. By
neglecting the contribution from matter fluids, that is, we are
interested in the vacuum $F(R)$ gravity case, we obtain the
following equations,
\begin{equation}
\label{cr12} 0= \frac{d^2 P(\phi)}{d\phi^2}
 - H \left( t=\phi \right) \frac{d P(\phi)}{d\phi}
+ 2 H'\left( t = \phi \right) P (\phi)\, , \quad Q(\phi) = - 6
H\left(t=\phi\right) \frac{d P}{d\phi} - 6 H \left( t=\phi \right)^2
P(\phi)\, ,
\end{equation}
or equivalently,
\begin{equation}
\label{cr12b} 0= \frac{d^2 P(\phi)}{d\phi^2} + M \tanh \left( \beta
M \phi \right) \frac{d P(\phi)}{d\phi} + \frac{2 \beta M^2}{\cosh^2
\left( \beta M \phi \right)} P (\phi)\, .
\end{equation}
If we assume that,
\begin{equation}
\label{cr13} P(\phi) \propto \sinh^\xi \left( \beta M \phi \right)
\cosh^\eta \left( \beta M \phi \right) \, ,
\end{equation}
where $\xi$ and $\eta$ are constant parameters, Eq.~(\ref{cr12b})
yields the following three algebraic equations,
\begin{equation}
\label{cr14} 0= \xi ( \xi - 1 ) \, , \quad 0=\left( \beta \eta -
\beta - 1 \right) \eta \, , \quad 0= \beta \left( \xi \left( \eta +
1 \right) + \eta \left( \xi + 1 \right) \right) + \xi + 2 \, .
\end{equation}
For general $\beta$, the above equations have no solution but if we
choose $\beta$ to take specific values, we find the following
solutions,
\begin{equation}
\label{cr15} \left( \xi,\eta, \beta \right) = \left( 0, \frac{2}{3},
-3 \right) , \ \left(1, 0, -2 \right) , \ \left( 1, -3, -
\frac{1}{4} \right) \, .
\end{equation}
In order to consider the more general case, we define a new variable
$y$ as follows,
\begin{equation}
\label{cr16} y=\frac{1}{\cosh^2 \left( \beta M \phi \right)}\, .
\end{equation}
Then by using the following relations,
\begin{align}
\label{cr17} \frac{d}{d\phi} =& - \frac{2\beta M \sinh \left( \beta
M \phi \right)} {\cosh^3 \left( \beta M \phi \right)}\frac{d}{dy}\,
, \nn \frac{d^2}{d\phi^2} =& \frac{4\beta^2 M^2 \sinh^2 \left( \beta
M \phi \right)} {\cosh^6 \left( \beta M \phi
\right)}\frac{d^2}{dy^2} + \beta^2 M^2 \left( \frac{4}{\cosh^2
\left( \beta M \phi \right)}
 - \frac{6}{\cosh^4 \left( \beta M \phi \right)} \right) \frac{d}{dy} \, ,
\end{align}
we can rewrite Eq.~(\ref{cr12b}) as follows,
\begin{equation}
\label{cr18} 0 = y \left( 1 - y \right) \frac{d^2P}{dy^2} + \left( 1
- \frac{1}{2\beta}
 - \left( \frac{3}{2} - \frac{1}{2\beta} \right) y \right)
\frac{dP}{dy} + \frac{1}{2\beta} P\, ,
\end{equation}
which is nothing but the hypergeometric differential equation and
the solutions are given by the hypergeometric functions as follows,
\begin{equation}
\label{cr19} P(\phi) = C_1 F\left( \alpha_+, \alpha_- ; \gamma ; y
\right) + C_2 y^{1-\gamma} F \left( \alpha_+ -\gamma + 1, \alpha_- -
\gamma + 1 ; 2 - \gamma ; y \right) \, .
\end{equation}
Here $C_1$ and $C_2$ are constants, which can depend on $\beta$, and
also
\begin{equation}
\label{cr20} \alpha_\pm = \frac{ 1 - \frac{1}{\beta} \pm \sqrt{
\left( 1 - \frac{1}{\beta} \right)^2 + \frac{8}{\beta} }}{4}\, ,
\quad \gamma = 1 - \frac{1}{2\beta}\, .
\end{equation}
Since the following holds true,
\begin{equation}
\label{cr20b} \frac{d F\left( \alpha_+, \alpha_- ; \gamma ; y
\right)}{dy} = \frac{\alpha_+ \alpha_-}{\gamma} F\left( \alpha_++1,
\alpha_-+1 ; \gamma+1 ; y \right)\, ,
\end{equation}
by using Eqs.~(\ref{cr12}) and (\ref{cr16}), we find,
\begin{align}
\label{cr20c} Q(\phi) =& C_1 \left( - 12 \beta M^2 y \left( 1 - y
\right) \frac{\alpha_+ \alpha_-}{\gamma} F\left( \alpha_++1,
\alpha_-+1 ; \gamma+1 ; y \right)
 - 6M^2 \left( 1 - y \right) F\left( \alpha_+, \alpha_- ; \gamma ; y \right)
\right) \nn & +  C_2 \Biggl( - 12\beta M^2 y \left( 1 - y \right)
\Bigl( \left( 1 - \gamma\right) y^{-\gamma} F \left( \alpha_+
-\gamma + 1, \alpha_- - \gamma + 1 ; 2 - \gamma ; y \right) \nn & +
\frac{ \left( \alpha_+ -\gamma + 1 \right) \left(\alpha_- - \gamma +
1\right)} {2 - \gamma} y^{1-\gamma} F \left( \alpha_+ -\gamma + 2,
\alpha_- - \gamma + 2 ; 3 - \gamma ; y \right) \Bigr) \nn & - 6M^2
\left( 1 - y \right) y^{1-\gamma} F \left( \alpha_+ -\gamma + 1,
\alpha_- - \gamma + 1 ; 2 - \gamma ; y \right) \Biggr) \, .
\end{align}
We are interested in the case that $\beta$ is of the order
$\mathcal{O}(1)$ and does not take large values. By taking this
limit, we find that,
\begin{equation}
\label{cr21} \alpha_+ \sim 1 \, , \quad \alpha_- \sim -
\frac{1}{2\beta}  - \frac{1}{2} \, , \quad \gamma \sim 1 -
\frac{1}{2\beta} \, ,
\end{equation}
and therefore we get,
\begin{align}
\label{cr22} P(\phi) =& C_1 F \left(1, - \frac{1}{2\beta};  -
\frac{1}{2\beta}; y \right) + C_2 y^{\frac{1}{2\beta}} F\left(
\frac{1}{2\beta}, \frac{1}{2}; \frac{1}{2\beta}; y \right) \nn = &
C_1 \frac{1}{1 - y} + C_2 \frac{y^{\frac{1}{2\beta}}}{\sqrt{1 - y}}
= C_1 \frac{1}{\tanh^2\left( \beta M \phi \right)} + C_2
\cosh^{-\frac{1}{\beta}}\left( \beta M \phi \right) \tanh\left(
\beta M \phi \right) \, .
\end{align}
Accordingly we find,
\begin{equation}
\label{cr23} Q(\phi) \sim C_1 \left( - \frac{12 \beta M^2}{
\sinh^2\left( \beta M \phi \right)} - 6 M^2 \right)
 - 6 C_2 \beta M^2
\cosh^{-\frac{1}{\beta}-2}\left( \beta M \phi \right) \tanh\left(
\beta M \phi \right) \, .
\end{equation}
We now consider the conformal transformation,
\begin{equation}
\label{cr24} g_{\mu\nu} \to \e^{\frac{\varphi}{\sqrt{3}}}
g_{\mu\nu}\, , \quad \frac{\varphi}{\sqrt{3}} = - \ln P(\phi)\, .
\end{equation}
and we rewrite the action (\ref{cr11}) as follows,
\begin{equation}
\label{cr25} S= \frac{1}{2\kappa^2} \int d^4 x \sqrt{-g} \left( R
 - \frac{1}{2} \partial^\mu \varphi \partial_\mu \varphi - V(\varphi) \right)\, ,
\quad V(\varphi) = - \frac{Q\left( \phi\left(\varphi\right) \right)
} {P\left( \phi\left(\varphi\right) \right)^2} \, .
\end{equation}
The time evolution appearing in Eq.~(\ref{cr10}) can be realized by
the Einstein frame scalar-tensor theory. The potential in the scalar
tensor theory in Eq.~(\ref{cr25}) is different from that in
\cite{Motohashi:2014ppa, Motohashi:2017aob}. Then there might be a
difference in the resulting spectral index of primordial curvature
perturbations $n_s$ and the scalar-to-tensor ratio $r$.

We may define the slow-roll parameters in the Einstein frame as
follows,
\begin{align}
\label{cr26} \epsilon \equiv& \frac{1}{2} \left(
\frac{V'(\varphi)}{V(\varphi)} \right)^2 = \frac{1}{6} \left(
\frac{P(\phi) Q'(\phi)}{P'(\phi) Q(\phi)} - 2  \right)^2 \, , \nn
\eta \equiv& \frac{V''(\varphi)}{V(\varphi)} = \frac{1}{3} \left(
\frac{Q''(\phi) P(\phi)^2}{Q(\phi)P'(\phi)^2}
 - 3 \frac{Q'(\phi) P(\phi)}{Q(\phi) P'(\phi) }
 - \frac{Q'(\phi) P(\phi)^2 P''(\phi)}{Q(\phi) P'(\phi)^3} + 4 \right) \, .
\end{align}
By using the slow-roll indexes $\epsilon$ and $\eta$, we can express
the observational indices $n_s$ and $r$ as follows,
\begin{equation}
\label{cr27} n_s -1 = - 6\epsilon + 2 \eta\, , \quad r=16\epsilon\,
.
\end{equation}
Then by setting $C_2=0$ in Eqs.~(\ref{cr22}) and (\ref{cr23}), in
the limit $\beta\to 0$, we get,
\begin{equation}
\label{cr28} \epsilon \sim \frac{2}{3}\, , \quad \eta \sim
\frac{4}{3} \, , \quad n_s - 1 \sim - \frac{4}{3}\, , \quad r \sim
\frac{32}{3}\, ,
\end{equation}
which seems too large compared with the observational data. On the
other hand, by setting $C_1=0$ in Eqs.~(\ref{cr22}) and
(\ref{cr23}), we obtain,
\begin{align}
\label{cr29} & \epsilon \sim \frac{1}{6} \left( \frac{1}{\cosh
\left( \beta M \phi \right) }
 - 2 \right)^2 \, , \quad
\eta \sim \frac{1}{3} \left( \cosh \left( \beta M \phi \right)
 - \frac{2}{\cosh \left( \beta M \phi \right)} + 4 \right) \, , \nn
& n_s - 1 \sim - \frac{1}{\cosh^2 \left( \beta M \phi \right) } +
\frac{4}{3 \cosh \left( \beta M \phi \right) } + \frac{8}{3} +
\frac{2}{3} \cosh \left( \beta M \phi \right) \, , \quad r \sim
\frac{8}{3} \left( \frac{1}{\cosh \left( \beta M \phi \right) }
 - 2 \right)^2 \, .
\end{align}
We should note that both the models obtained in the cases $C_2=0$
and $C_1=0$ yield the same background cosmological evolution, but
the resulting observational indices are different from each other.
It is possible though to adjust the coefficients $C_1$ and $C_2$ in
such a way so that these depend on $ \beta$. In this way we may
obtain a variety of $n_s$ and $r$.

It is also possible to realize cosmological models that in the
scalar-tensor frame allow transitions between constant-roll eras,
which correspond to different parameters $\beta$. Since we
identified the scalar field $\phi$ with the cosmological time, the
model describing the transition between constant-roll eras can be
realized by allowing the parameter $\beta$ to depend on $\phi$, as
follows,
\begin{align}
\label{cr30} P(\phi) =& C_1\left( \beta(\phi) \right) F\left(
\alpha_+(\phi), \alpha_-(\phi) ; \gamma(\phi) ; y \right) +
C_2\left( \beta(\phi) \right) y^{1-\gamma(\phi)} F \left(
\alpha_+(\phi) -\gamma(\phi) + 1, \alpha_-(\phi) - \gamma(\phi) + 1
; 2 - \gamma(\phi) ; y \right) \, , \nn Q(\phi) =& C_1\Biggl( - 12
\beta(\phi) \beta(\phi) M^2 y \left( 1 - y \right)
\frac{\alpha_+(\phi) \alpha_-(\phi)}{\gamma(\phi)} F\left(
\alpha_+(\phi)+1, \alpha_-(\phi)+1 ; \gamma(\phi)+1 ; y \right) \nn
& - 6M^2 \left( 1 - y \right)
 F\left( \alpha_+(\phi), \alpha_-(\phi) ; \gamma(\phi) ; y \right) \Biggr) \nn
& + C_2 \Biggl( - 12 \beta(\phi) M^2 y \left( 1 - y \right) \Bigl(
\left( 1 - \gamma(\phi)\right) y^{-\gamma(\phi)} F \left(
\alpha_+(\phi) -\gamma(\phi) + 1, \alpha_-(\phi) - \gamma(\phi) + 1
; 2 - \gamma(\phi) ; y \right)  \nn & + \frac{ \left( \alpha_+(\phi)
-\gamma(\phi) + 1 \right) \left(\alpha_-(\phi) - \gamma(\phi) +
1\right)} {2 - \gamma(\phi)} y^{1-\gamma(\phi)} F \left(
\alpha_+(\phi) -\gamma(\phi) + 2, \alpha_-(\phi) - \gamma(\phi) + 2
; 3 - \gamma(\phi) ; y \right) \Bigr) \nn & - 6M^2 \left( 1 - y
\right) y^{1-\gamma(\phi)} F \left( \alpha_+(\phi) -\gamma(\phi) +
1, \alpha_-(\phi) - \gamma(\phi) + 1 ; 2 - \gamma(\phi) ; y \right)
\Biggr) \, , \nn \alpha_\pm(\phi) =& \frac{ 1 -
\frac{1}{\beta(\phi)} \pm \sqrt{ \left( 1 - \frac{1}{\beta(\phi)}
\right)^2 + \frac{8}{\beta(\phi)} }}{4}\, , \quad \gamma = 1 -
\frac{1}{2\beta(\phi)}\, .
\end{align}
Then, we may choose,
\begin{equation}
\label{cr31} \beta(\phi) = \beta_1 + \left( \beta_2 - \beta_1
\right) \tanh \left( \tilde M \left( \phi - t_0 \right) \right)\, ,
\end{equation}
where, $\beta_1$, $\beta_2$, $\tilde M$, and $t_0$ are constants.
Then in the case $\phi=t \ll t_0$, the constant-roll era with
$\beta\to \beta_1$ is realized, and if $\phi=t \gg t_0$, the
constant-roll era with $\beta\to \beta_2$ is realized. The
corresponding $F(R)$ gravities can easily be found, but we omit the
result for brevity, since it is too lengthy to be presented here.

We should note that even if we replace $\beta$ in (\ref{cr12}) with
$\beta(\phi)$, which is a function of the scalar field $\phi$, the
expressions in (\ref{cr30}) are not the solution of the obtained
equation. As we see in Eq.~(\ref{cr31}), however, we have chosen so
that $\beta(\phi)$ becomes a constant in the limits of $\phi \to \pm
\infty$. In the limits, the scalar field $\phi$ can be identified
with the time coordinate although we cannot identify it as the time
coordinate for finite $\phi$. In the limits,the expressions in
(\ref{cr30}) satisfy (\ref{cr12}) asymptotically and the model given
in (\ref{cr30}) really connects two different $\beta$'s.

Another interesting scenario which realizes a transition between
constant-roll eras can be shown \cite{Odintsov:2017qpp}, that has
the following canonical scalar field potential,
\begin{equation}
\label{potentialcase1} V(\varphi)=2 \beta ^2 M_p^2
\e^{\frac{\sqrt{2} \varphi }{M_p}} +6 \beta  \delta  M_p^2
\e^{\frac{\varphi }{\sqrt{2} M_p}} +3 \delta ^2 M_p^2\, ,
\end{equation}
in which case the resulting Hubble evolution is,
\begin{equation}
\label{hubbleevolutioncctrans} H(t)=\delta +\frac{1}{t}\, ,
\end{equation}
and also the transition is ensured by the condition that the second
slow-roll index $\eta$ satisfies,
\begin{equation}
\label{basiccondition}
\eta=-\frac{\ddot{\varphi}}{2H\dot{\varphi}}=\frac{\beta  \exp
(\lambda  \varphi)}{\delta +\beta  \exp (\lambda  \varphi)}\, .
\end{equation}
The qualitative study of this constant-roll to constant-roll
transition scenario can be found in Ref.~\cite{Odintsov:2017qpp}.
Here we shall investigate how the evolution
(\ref{hubbleevolutioncctrans}) can be realized by using a vacuum
$F(R)$ gravity and we shall mainly be interested in the Einstein
frame theory corresponding to the $F(R)$ gravity. As we shall show,
the resulting Einstein frame potential is not identical to the one
appearing in Eq.~(\ref{potentialcase1}). By using the formalism we
presented earlier in this section, it can easily be found that the
function $P(\phi)$ is equal to
\begin{equation}
\label{pf1} P(\phi)=c_1 \phi ^{1+\sqrt{3}} U\left(1+\sqrt{3},1+2
\sqrt{3},\delta \phi \right)+c_2 \phi ^{1+\sqrt{3}}
L_{-1-\sqrt{3}}^{2 \sqrt{3}}(\delta  \phi )\, ,
\end{equation}
where $U(x,y,z)$ is the confluent hypergeometric function and
$L_n^m(z)$ is the generalized Laguerre polynomial, and $c_i$,
$i=1,2$ are integration constants. By using Eq.~(\ref{cr12}), the
function $Q(\phi)$ can also be found and it is equal to,
\begin{align}
\label{qfasx} Q(\phi) = & -6 \phi ^{\sqrt{3}-1} \delta  \phi +
\left( c_1 \left(\delta \phi +\sqrt{3}+2\right)
U\left(1+\sqrt{3},1+2 \sqrt{3},\delta  \phi \right) \right. \nn &
-\left(1+\sqrt{3}\right) c_1 \delta \phi U\left(2+\sqrt{3},2+2
\sqrt{3},\delta  \phi \right) \nn & \left. + c_2 \left(\left(\delta
\phi +\sqrt{3}+2\right) L_{-1-\sqrt{3}}^{2 \sqrt{3}}(\delta  \phi
)-\delta  \phi L_{-2-\sqrt{3}}^{1+2 \sqrt{3}}(\delta  \phi )\right)
\right)\, .
\end{align}
Having the functions $P(\phi)$ and $Q(\phi)$ at hand, we can use
Eqs.~(\ref{cr24}) and (\ref{cr25}) to find the potential
$V(\varphi)$. However, the function $\phi (\varphi) $ is not so easy
to find. Therefore, in order to have an idea how the Einstein frame
potential looks like, we perform an asymptotic expansion of the
function $P(\phi)$ for small values of $\delta$, and we get at
leading order,
\begin{equation}
\label{pf2} P(\phi )\simeq \gamma +\delta ^{-2 \sqrt{3}} \lambda
\phi ^{-2 \sqrt{3}}\, ,
\end{equation}
where the constant parameters $\gamma$ and $\lambda$ are,
\begin{equation}
\label{parmeneters} \gamma=\binom{-1+\sqrt{3}}{2 \sqrt{3}} c_2\,
,\quad \lambda =\frac{c_1 \Gamma \left(2 \sqrt{3}\right)}{\Gamma
\left(1+\sqrt{3}\right)}\, .
\end{equation}
Then, the function $\phi (\varphi)$ reads,
\begin{equation}
\label{asxet} \phi (\varphi)=\frac{\left(\e^{\frac{\varphi
}{\sqrt{3}}}-\gamma \right)^{-\frac{1}{2 \sqrt{3}}}}{\delta  \lambda
^{-\frac{1}{2 \sqrt{3}}}}\, ,
\end{equation}
hence the potential $V(\varphi)$ can be found by combining
Eqs.~(\ref{cr12}), (\ref{cr24}), (\ref{cr25}), and (\ref{pf2}), and
it reads,
\begin{align}
\label{varphippotential} V(\varphi)\simeq& -\frac{6 \delta ^2
\lambda ^{-\frac{1}{\sqrt{3}}} \left(\left(\e^{\frac{\varphi
}{\sqrt{3}}}-\gamma \right)^{\frac{1}{2 \sqrt{3}}}+\lambda
^{\frac{1}{2 \sqrt{3}}}\right) \left(-\gamma
\left(\left(\e^{\frac{\varphi }{\sqrt{3}}}-\gamma
\right)^{\frac{1}{2 \sqrt{3}}}+\lambda ^{\frac{1}{2
\sqrt{3}}}\right)\right)}{\left(\e^{\frac{\varphi
}{\sqrt{3}}}-\gamma \right)^2} \nn & -\frac{6 \delta ^2 \lambda
^{1-\frac{1}{2 \sqrt{3}}} \left(\lambda ^{\frac{1}{2 \sqrt{3}}}
\left(-\left(\e^{\frac{\varphi }{\sqrt{3}}}-\gamma
\right)^{-\frac{1}{2 \sqrt{3}}}\right)+2
\sqrt{3}-1\right)}{\left(\e^{\frac{\varphi }{\sqrt{3}}}-\gamma
\right)^2}\, .
\end{align}
By comparing Eqs.~(\ref{potentialcase1}) and
(\ref{varphippotential}), it can be seen that the scalar potentials
are totally different, and the same applies even if we take the
small $\delta$ limit of the potential (\ref{potentialcase1}). It is
conceivable then that the potential (\ref{varphippotential})
describes an entirely different cosmological evolution, and also it
is not certain that the constant-roll condition still holds true.
However we refer from discussing in detail these issues here.

\section{The Constant-roll Inflation Condition with $F(R)$ Gravity
\label{SecIII}}

In the previous section our approach to the constant-roll
inflationary scenarios was in some way indirect since we realized
the Hubble rate corresponding to the scalar-tensor frame theories,
by using $F(R)$ gravity. In the present section we shall use the
constant-roll condition directly in the $F(R)$ gravity theory and we
shall discuss the new qualitative features that the constant-roll
condition imposes in the $F(R)$ gravity inflationary phenomenology.
We will be interested in two vacuum $F(R)$ gravity models, the $R^2$
model and a power-law $F(R)$ gravity model.

\subsection{General Formalism of the Constant-Roll Inflation with $F(R)$ Gravity \label{SecIIIA}}

First let us recall here that the constant-roll inflation condition
has the following form in terms of the Hubble rate,
\begin{equation}
\label{constantrollcondition} \frac{\ddot{H}}{2H\dot{H}}\simeq
\beta\, ,
\end{equation}
where $\beta$ is a real parameter. In this section we shall present
the theoretical framework of the $F(R)$ gravity inflation, with the
assumption that the above condition holds true. We shall assume that
the physical evolution is controlled by a vacuum $F(R)$ gravity with
action,
\begin{equation}
\label{JGRG7} S_{F(R)}= \int d^4 x \sqrt{-g} \left(
\frac{F(R)}{2\kappa^2} \right)\, ,
\end{equation}
where $g$ stands for the determinant of the background metric, which
shall be assumed to be a flat FRW metric. By varying the action
(\ref{JGRG7}) with respect to the metric, we obtain the following
equations of motion,
\begin{align}
\label{eqnmotion1}
3F_RH^2=& \frac{F_RR-F}{2}-3H\dot{F}_R \, , \\
\label{eqnmotion2} -2F_R\dot{H}=& \ddot{F}-H\dot{F} \, ,
\end{align}
where $F_R=\frac{\partial F}{\partial R}$ and the ``dot'' indicates
differentiation with respect to the cosmic time. The inflationary
dynamics in the context of modified gravity are well described in
Refs.~\cite{Noh:2001ia,Hwang:2001qk,Hwang:2001pu}, see also
Refs.~\cite{Nojiri:2016vhu,Odintsov:2016plw,Odintsov:2015gba} for
some recent works. The inflationary dynamics are perfectly described
by the following inflationary indices, which for a general $F(R)$
gravity are equal to,
\begin{equation}
\label{slowrollgenerarlfrphi}
\epsilon_1=-\frac{\dot{H}}{H^2}\,,\quad \epsilon_2=0\, , \quad
\epsilon_3=\frac{\dot{F}_R}{2HF_R}\, ,\quad
\epsilon_4=\frac{\dot{E}}{2HE}\, ,
\end{equation}
where the function $E$ appearing in
Eq.~(\ref{slowrollgenerarlfrphi}) stands for,
\begin{equation}
\label{epsilonfnction} E=\frac{3\dot{F}_R^2}{2\kappa^2}\, .
\end{equation}
Another useful quantity which we shall now introduce is the function
$Q_s$, which is,
\begin{equation}
\label{qsfunction} Q_s=\frac{E}{F_RH^2(1+\epsilon_3)^2}\, ,
\end{equation}
which we shall use later on when we calculate the scalar-to-tensor
ratio.

We will be mainly interested in the calculation of the spectral
index of the primordial curvature perturbations $n_s$ and of the
scalar-to-tensor ratio in the context of pure $F(R)$ gravity. With
regard to the spectral index, in the case that
$\dot{\epsilon}_i\simeq 0$, it is equal to
\cite{Noh:2001ia,Hwang:2001qk,Hwang:2001pu},
\begin{equation}
\label{spectralindex1} n_s=4-2\nu_s\, ,
\end{equation}
where $\nu_s$ is equal to,
\begin{equation}
\label{nus}
\nu_s=\sqrt{\frac{1}{4}+\frac{(1+\epsilon_1-\epsilon_3+\epsilon_4)
(2-\epsilon_3+\epsilon_4)}{(1-\epsilon_1)^2}}\, .
\end{equation}
In the case that $\epsilon_i\ll 1$, the spectral index can be
approximated as follows,
\begin{equation}
\label{spectralindex2} n_s\simeq
1-4\epsilon_1+2\epsilon_3-2\epsilon_4\, .
\end{equation}
Now we turn our focus on the scalar-to-tensor ratio, which in the
case of $F(R)$ gravity is defined as follows,
\begin{equation}
\label{scalartotensor1} r=\frac{8\kappa^2Q_s}{F_R}\, ,
\end{equation}
where $Q_s$ is defined in Eq.~(\ref{qsfunction}). After some
algebra, it can be shown that in the case of $F(R)$ gravity, the
scalar-to-tensor ratio reads,
\begin{equation}
\label{scalartotensor2} r=\frac{48\epsilon_3^2}{(1+\epsilon_3)^2}\,
.
\end{equation}

At this point we shall investigate how the inflationary indices and
also the observational indices are affected by the constant-roll
condition. It can be shown after some algebra that the inflationary
indices (\ref{slowrollgenerarlfrphi}) become,
\begin{equation}
\label{frgravityconstantroll}
\epsilon_1=-\frac{\dot{H}}{H^2}\,,\quad \epsilon_2=0\, , \quad
\epsilon_3=\frac{\dot{F}_{RR}}{2HF_R}\left(
24H\dot{H}+\ddot{H}\right)\, ,\quad
\epsilon_4=\frac{F_{RRR}}{HF_R}\dot{R}+\frac{\ddot{R}}{H\dot{R}}\, ,
\end{equation}
where $F_{RR}=\frac{\partial^2 F}{\partial R^2}$ and
$F_{RRR}=\frac{\partial^3 F}{\partial R^3}$. By imposing the
condition (\ref{constantrollcondition}), it can be shown that the
following approximations hold true,
\begin{align}
\label{approximations1}
\dot{R}=& 12H\dot{H}(\beta+2)\, , \\
\label{approximations2}
\ddot{R}=&12H(\beta+2)(\dot{H}^2+\ddot{H}H)\, .
\end{align}
Depending on the functional form of the $F(R)$ gravity, the
inflationary indices $\epsilon_i$, $i=1,\cdots, 4$ and the
corresponding observational indices $n_s$ and $r$, can take various
forms, so this is the subject of this section. We need to note
however that the in the limit $\beta\to 0$, the constant-roll and
the slow-roll expressions should definitely coincide.

Before we proceed to the models we would like to note that we shall
assume that the parameter $\epsilon_1$ still satisfies
$\epsilon_1\ll 1$, an assumption also made in \cite{Martin:2012pe}.

\subsection{The Starobinsky $R^2$ Model with Constant-Roll Inflation
Condition \label{SecIIIB}}

\begin{figure}[h]
\centering
\includegraphics[width=18pc]{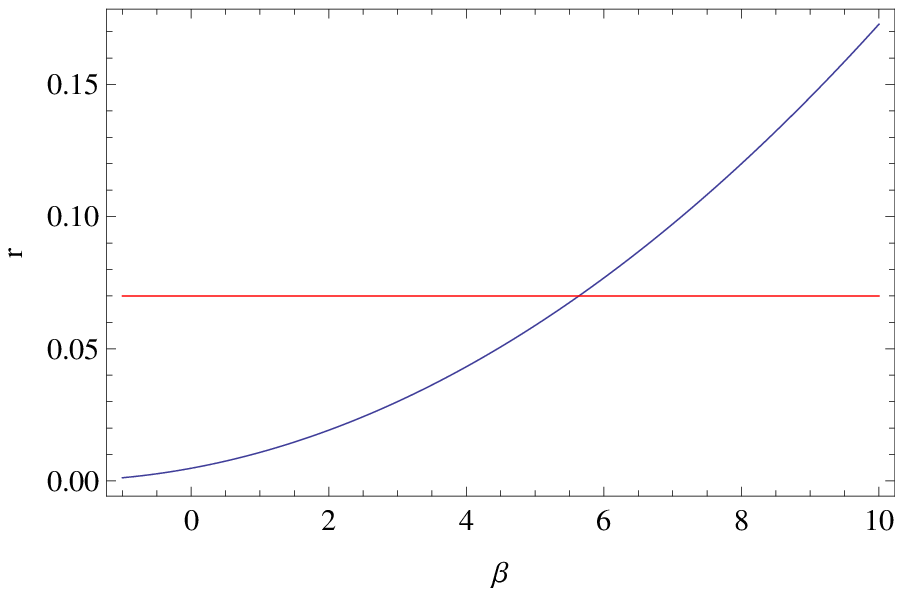}
\includegraphics[width=18pc]{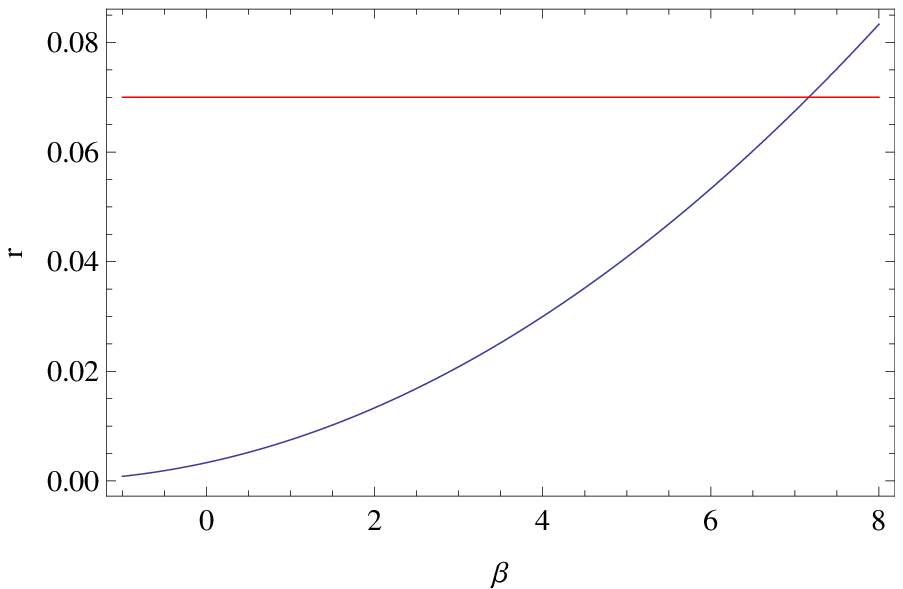}
\caption{The scalar-to-tensor ratio $r=\frac{3 (\beta +2)^2}{N^2}$,
as a function of the parameter $\beta$, for $N=50$ (blue curve, left
plot) and for $N=60$ (blue curve, right plot). In both plots, the
red curve corresponds to the 2015-BICEP2/Keck-Array upper bound
constraint $r=0.07$. }\label{plot2}
\end{figure}

In this section we shall investigate the theoretical implications of
the constant-roll condition on the inflationary phenomenology of the
$R^2$ inflation model, in which case the $F(R)$ gravity is of the
form \cite{Starobinsky:1980te},
\begin{equation}
\label{r2inflation} F(R)=R+\frac{1}{36H_i}R^2\, ,
\end{equation}
where $H_i$ is a phenomenological parameter which has dimensions of
mass$^2$ and it is assumed to be quite large $H_i\gg 1$. For the
$R^2$ model (\ref{r2inflation}) it can be easily shown that the
inflationary indices of Eq.~(\ref{frgravityconstantroll}) become,
\begin{equation}
\label{frgravityconstantroll1r2}
\epsilon_1=-\frac{\dot{H}}{H^2}\,,\quad \epsilon_2=0\, , \quad
\epsilon_3=-\frac{12\beta+24}{24}\epsilon_1\, ,\quad
\epsilon_4=-3\epsilon_1+\frac{\dot{\epsilon}_1}{H\epsilon_1}\, .
\end{equation}
Due to the fact that we assumed $\epsilon_1\ll 1$, we approximately
have $\dot{\epsilon}_i\simeq 0$, therefore we can use the relation
(\ref{spectralindex2}) for the spectral index. In addition, the
scalar-to-tensor ratio can be calculated by using
Eq.~(\ref{scalartotensor2}). In order to calculate the inflationary
indices, we need to have an approximate expression for the Hubble
rate during the era for which $\epsilon_1\ll 1$, with the
constant-roll condition (\ref{constantrollcondition}) approximately
holding true. Hence we will solve the equations of motion
(\ref{eqnmotion1}) and (\ref{eqnmotion2}), for the $F(R)$ gravity
(\ref{r2inflation}) by using the constant-roll condition
$\ddot{H}\sim 2\beta H\dot{H}$. During the era for which
$\epsilon_1\ll 1$, or equivalently when the condition $\dot{H}\ll
H^2$ holds true, the equations of motion (\ref{eqnmotion1}) and
(\ref{eqnmotion2}) become,
\begin{equation}
\label{frweqnsr2} \ddot{H}-\frac{\dot{H}^2}{2H}+3H_iH=-3H\dot{H}\,
,\quad \ddot{R}+3H\dot{R}+6H_iR=0\, .
\end{equation}
By using the constant-roll condition $\ddot{H}\sim 2\beta H\dot{H}$,
the first differential equation in (\ref{frweqnsr2}) becomes,
\begin{equation}
\label{rsquarebasic} \dot{H}H \left( 2\beta+\frac{\epsilon_1}{2}+3
\right)\dot{H}=-3H_i\, ,
\end{equation}
and since $\epsilon_1\ll 1$, by eliminating the $\epsilon_1$
dependence in the differential equation (\ref{rsquarebasic}), we
find the approximate solution,
\begin{equation}
\label{hubblersquare} H(t)=H_0-H_I(t-t_k)\, ,
\end{equation}
where $H_0$ is arithmetically of the order $\mathcal{O}(H_i)$, but
with different dimensions, and also the parameter $H_I$ is,
\begin{equation}
\label{hI} H_I=\frac{3H_i}{2\beta+3}\, .
\end{equation}
In addition, the time instance $t=t_k$ is the horizon crossing time
instance. The cosmological evolution (\ref{hubblersquare}) is a
quasi-de Sitter evolution, a bit different from the ordinary $R^2$
model, however in the limit $\beta\to 0$, these two coincide. In the
following we shall be interested in finding the inflationary
dynamics of the approximate quasi-de Sitter solution
(\ref{hubblersquare}). The inflationary era will eventually stop if
the first slow-roll index becomes of the order $\epsilon_1\simeq
\mathcal{O}(1)$, so by assuming that this occurs at a time instance
$t_f$, with $H(t_f)=H_f$, the condition $\epsilon_1(t_f)\simeq 1$
yields, $H_f\simeq \sqrt{H_I}$. Then we obtain,
\begin{equation}
\label{finalghf} H_f-H_0\simeq -H_I (t_f-t_k)\, ,
\end{equation}
and by substituting $H_f$ we get,
\begin{equation}
\label{timerelation}
t_f-t_k=\frac{H_0}{H_I}-\frac{\sqrt{H_I}}{H_I}\, .
\end{equation}
During the era $\epsilon_1\ll 1$, the parameters $H_0,H_I$ take
large values, so by omitting the last term in (\ref{timerelation})
we obtain,
\begin{equation}
\label{timerelation1} t_f-t_k\simeq \frac{H_0}{H_I}\, .
\end{equation}
It is worth invoking the $e$-foldings number $N$ in the calculation,
which is defined as,
\begin{equation}
\label{efold1} N=\int_{t_k}^{t_f}H(t)d t\, .
\end{equation}
By substituting the Hubble rate (\ref{hubblersquare}) in the
$e$-foldings number (\ref{efold1}) we get,
\begin{equation}
\label{nefold1} N=H_0(t_f-t_k)-\frac{H_I(t_f-t_k)^2}{2}\, ,
\end{equation}
so by substituting Eq.~(\ref{timerelation1}), we obtain,
\begin{equation}
\label{nfinal1} N=\frac{H_0^2}{2H_I}\, .
\end{equation}
Hence at leading order we have approximately,
\begin{equation}
\label{leadingtf} t_f-t_k\simeq \frac{2N}{H_0}\, .
\end{equation}
Having the above relations at hand we can calculate the inflationary
indices (\ref{frgravityconstantroll1r2}) and the observational
indices (\ref{spectralindex2}) and (\ref{scalartotensor2}). Hence,
by calculating the spectral index we obtain at leading order for
large-$N$,
\begin{equation}
\label{spectralindexresult1} n_s\simeq 1-\frac{\beta +4}{2 N}\, .
\end{equation}
Accordingly, the scalar-to-tensor ratio at leading order is,
\begin{equation}
\label{rsleadingorderr2} r\simeq \frac{3 (\beta +2)^2}{N^2}\, .
\end{equation}
For the ordinary $R^2$ model, the observational indices are,
\begin{equation}
\label{jordanframeattract} n_s\simeq 1-\frac{2}{N}\, ,\quad r\simeq
\frac{12}{N^2}\, ,
\end{equation}
and by comparing Eqs.~(\ref{spectralindexresult1}) and
(\ref{rsleadingorderr2}) with (\ref{jordanframeattract}), it can
easily be seen that in the limit $\beta\to 0$, these coincide. Let
us now investigate the viability of the constant-roll $R^2$ model by
comparing the observational indices with the Planck data
\cite{Ade:2015lrj} and also with the BICEP2/Keck-Array data
\cite{Array:2015xqh}, for specific values of $N$ and $\beta$. The
constraints on the spectral index and on the scalar-to-tensor ratio
imposed by the Planck data, are as follows,
\begin{equation}
\label{planckdata} n_s=0.9644\pm 0.0049\, , \quad r<0.10\, ,
\end{equation}
and also the BICEP2/Keck-Array data \cite{Array:2015xqh} constraints
the scalar-to-tensor ratio even further, in the following way,
\begin{equation}
\label{scalartotensorbicep2} r<0.07\, ,
\end{equation}
at $95\%$ confidence level. Let us now investigate in some detail
the parameter space of the constant-roll Starobinsky model
quantified by the parameters $(N,\beta)$. An analysis immediately
reveals that the compatibility with the observational data can be
achieved for a wide range of values for the parameters, and we now
try to demonstrate this by using some illustrative plots. Firstly
let us comment that the spectral index is considered within the
observational constraints if it takes values in the interval
$n_s=[0.9595,0.9693]$, so we take this into account in our analysis.
\begin{figure}[h]
\centering
\includegraphics[width=18pc]{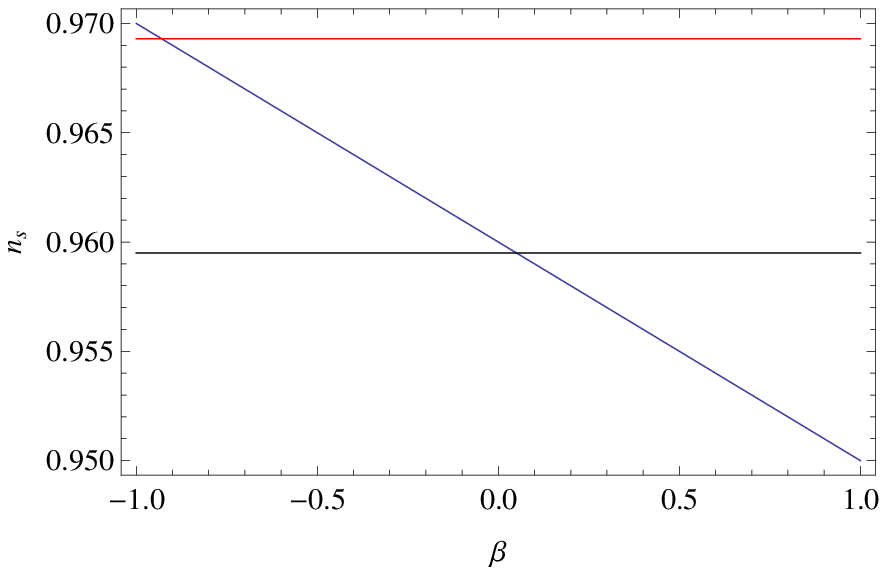}
\includegraphics[width=18pc]{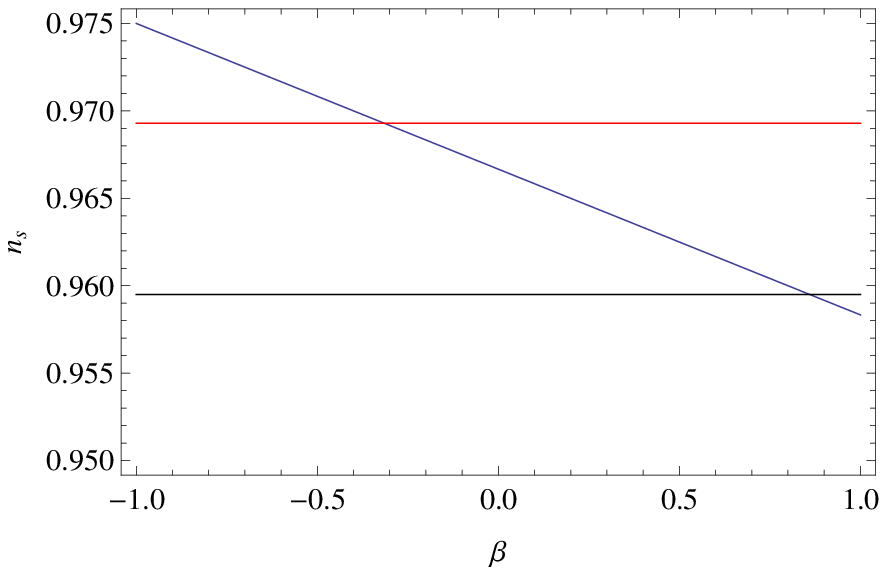}
\caption{The spectral index $n_s=1-\frac{\beta +4}{2 N}$ as a
function of the parameter $\beta$, for $N=50$ (blue curve, left
plot) and for $N=60$ (blue curve, right plot). The red line in both
plots corresponds to the upper bound of the 2015-Planck constraints
on the spectral index $n_s=0.9693$, while the black line in both
plots corresponds to the lower bound of the 2015-Planck data
$n_s=0.9595$.}\label{plot1}
\end{figure}
In Fig.~\ref{plot1} we plotted the $\beta$ dependence of the
spectral index for $N=50$ (left plot) and for $N=60$ (right plot).
In both plots, the upper (red) and lower (black) straight lines
correspond to the values $n_s=0.9693$ and $n_s=0.9595$ respectively.
As it can be seen, there is a large range of values for the
parameter $\beta$, for which the spectral index becomes compatible
with observations and this can be achieved for various values of the
$e$-foldings number. Let us here present some characteristic
examples, starting with the set of values $(N,\beta)=(45,-3)$, with
the $\beta=-3$ constant-roll scenario being known in the literature
as ultra-slow-roll scenario \cite{Martin:2012pe}. For
$(N,\beta)=(15,-3)$ we obtain,
\begin{equation}
\label{nsr} n_s=0.966667\, ,\quad r=0.0133333\, ,
\end{equation}
which are both within the Planck and BICEP2/Keck-Array data
constraints, however this scenario is not so appealing since the
$e$-foldings number is too small. Also for $(N,\beta)=(50,-0.6)$ we
obtain,
\begin{equation}
\label{nsr} n_s=0.9667\, ,\quad r=0.002352\, ,
\end{equation}
which again are both within the Planck and BICEP2/Keck-Array data
constraints. Another interesting constant-roll example corresponds
to the value $\beta=0.01$ which belongs to the models studied in
\cite{Motohashi:2017aob}, in which case for $N=60$ the observational
indices read,
\begin{equation}
\label{nsr} n_s=0.966583\, ,\quad r=0.00336675\, ,
\end{equation}
and these are observationally acceptable. From the analysis we
performed it seems that both the constant-roll models with negative
and positive $\beta$ produce viable results, at least for the
Starobinsky $R^2$ model in vacuum. Also in Fig.~\ref{plot2}, the
scalar-to-tensor ratio is plotted as a function of $\beta$ for
$N=50$ (left plot), and for $N=60$ (right plot) with the red line
indicating the BICEP2/Keck-Array constraint $r=0.07$. As it can be
seen, there is a large range of $\beta$ values for which the
resulting scalar-to-tensor ratio is in concordance with the
observational data.
Before closing this section we discuss two issues related to the
horizon crossing and the end of inflation. For the derivation of the
observational indices, we calculated the inflationary indices at the
horizon crossing, and hence the same applies for the observational
indices. While this is not necessarily the case in the scalar-tensor
description as it can be seen in Ref.~\cite{Martin:2012pe}, in our
case, since all the inflationary indices $\epsilon_i$ are quite
small when $\epsilon_1\ll 1$, the calculation can be performed at
the horizon crossing. Also we assumed that inflation ends when
$\epsilon_1\sim 1$, however this can be questionable. Traditionally,
when $\epsilon_1\sim 1$, the slow-roll era ends, and this is usually
identified with the end of the inflationary era. However, the actual
ending of inflation is caused by growing curvature perturbations
which render the final de Sitter attractor unstable. In our case the
final attractor is a quasi-de Sitter attractor and since we are
dealing with an $R^2$ model, the $R^2$ term is known to produce the
exit from inflation \cite{Bamba:2014jia}, due to growing curvature
perturbations. It is worth briefly discussing this issue at this
point, so let us consider how the perturbations from the de Sitter
solution grow as a function of the cosmic-time. This might
eventually impose some restrictions on the parameter $\beta$.
Consider the following perturbation of the de Sitter solution,
\begin{equation}
\label{perturbationfromdesitter} H(t)=H_0+\Delta H(t)\, ,
\end{equation}
so we substitute (\ref{perturbationfromdesitter}) in
Eq.~(\ref{eqnmotion1}), and by keeping linear terms of the function
$\Delta H(t)$, $\Delta \dot{H}(t)$ and also by using the
constant-roll condition (\ref{constantrollcondition}), we obtain the
following differential equation,
\begin{equation}
\label{evolutioneqntiondheta} 6 H_0^2 H_i+2 \beta  H_0^2 \Delta
\dot{H}(t)+6 H_0^2 \Delta \dot{H}(t)+12 H_0 H_i \Delta H(t)=0\, .
\end{equation}
The differential equation (\ref{evolutioneqntiondheta}) determines
the evolution of linear perturbations of the de Sitter solution, and
it can be solved analytically, with the solution being,
\begin{equation}
\label{eylogi} \Delta H(t)=C_1 \e^{-\frac{6 H_i t}{H_0 (\beta
+3)}}-\frac{H_0}{2}\, ,
\end{equation}
where $C_1$ is an integration constant. From the above solution,
since $H_0\gg 1$, in order to have growing perturbations, the
parameter $\beta$ must satisfy $\beta<-3$. However this restriction
leads to a rather questionable result, since for $n<-3$, the
$e$-foldings number must be much smaller than $n\sim 50-60$ in order
to obtain compatibility with the observational data. This means that
the constant-roll scenario lasts only $10-15$ $e$-foldings, which is
a relatively small period. This is possibly an indication that for
$\beta<-3$, the constant-roll scenario becomes quite unstable,
however this needs to be further investigated, since the approach we
adopted is based on keeping linear perturbation terms, hence our
results might be an artifact of the linear perturbation theory.

\subsection{Power-law $F(R)$ Gravity Model with Constant-Roll Inflation
Condition \label{SecIIIC}}

Now let us consider another interesting $F(R)$ gravity model, in
which case the $F(R)$ has the following form,
\begin{equation}
\label{r2inflation1} F(R)=\alpha R^n\, ,
\end{equation}
where the parameters $\alpha$ and $n$ are positive numbers. By
taking into account the constant-roll condition
(\ref{constantrollcondition}), the inflationary indices of
Eq.~(\ref{frgravityconstantroll}) take the following form,
\begin{equation}
\label{frgravityconstantroll1r2rnmodel}
\epsilon_1=-\frac{\dot{H}}{H^2}\,,\quad \epsilon_2=0\, , \quad
\epsilon_3=-(n-1) \left( 1+\frac{\beta}{2} \right)\epsilon_1\,
,\quad \epsilon_4=-\left( \left(n-2\right) \left(\beta+2 \right) +3
\right)\epsilon_1\, .
\end{equation}
It can be easily confirmed by looking the related literature that
the expressions above coincide with the ordinary power-law $F(R)$
gravity model when $\beta=0$. In order to proceed and further
demonstrate this coincidence for $\beta=0$, we will calculate the
approximate form of the Hubble rate by using the $F(R)$ gravity
equations of motion. Particularly, the differential equation
(\ref{eqnmotion1}) when $H^2\gg \dot{H}$, it takes the following
approximate form,
\begin{equation}
\label{rsquarebasic1}
6nH^2-12(n-1)H^2-6(n-1)\dot{H}+6n(n-1)\dot{H}(\beta+2)=0\, ,
\end{equation}
and by analytically solving this equation we obtain the following
solution,
\begin{equation}
\label{hubblersquare1} H(t)=\frac{(n-1) ((\beta +2) n-1)}{(2-n) t}\,
,
\end{equation}
which for $\beta=0$ coincides with the well known solution
corresponding to the ordinary slow-roll power-law $F(R)$ gravity
model.

\begin{figure}[h]
\centering
\includegraphics[width=18pc]{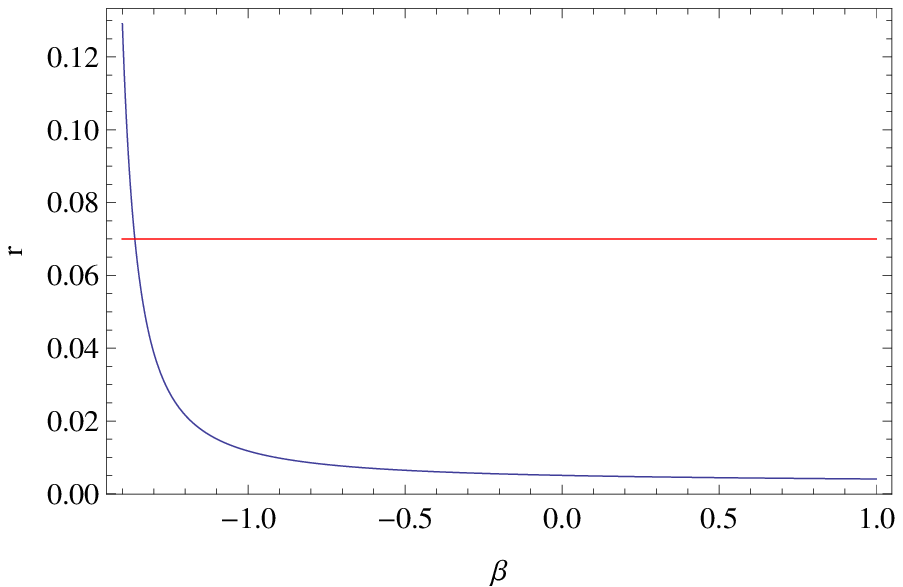}
\includegraphics[width=18pc]{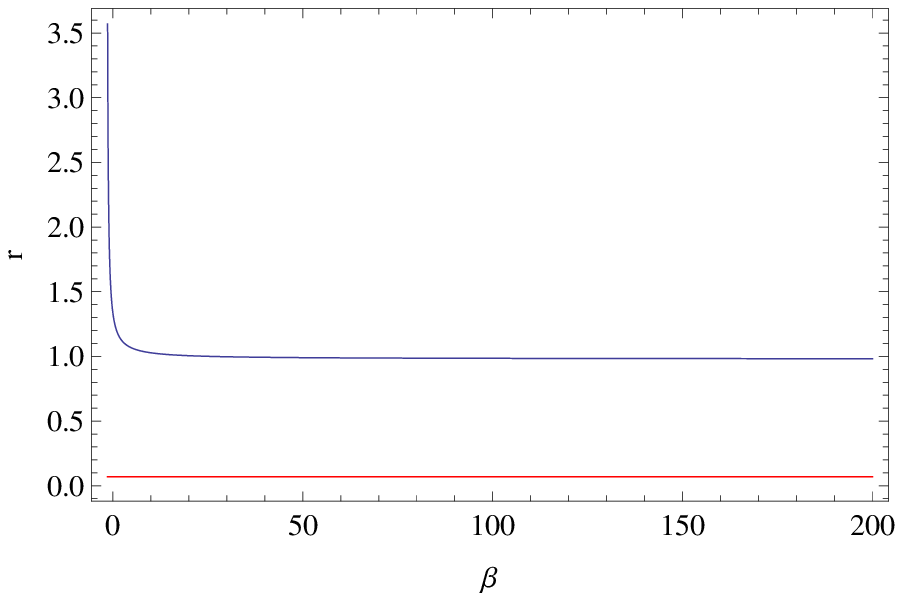}
\caption{The scalar-to-tensor ratio $r=\frac{48 (\beta +2)^2
(n-2)^2}{(3 (\beta +2) n-2 (\beta +3))^2}$, as a function of the
parameter $\beta$, for $n=1.97$ (blue curve, left plot) and for
$n=3$ (blue curve, right plot). In both plots, the red curve
corresponds to the 2015-BICEP2/Keck-Array upper bound constraint
$r=0.07$. }\label{plot4}
\end{figure}

We proceed to the calculation of the inflationary indices by taking
into account the solution (\ref{hubblersquare1}), and these become,
\begin{equation}
\label{frgravityconstantroll1r2rnmodel11111}
\epsilon_1=\frac{2-n}{(n-1) ((\beta +2) n-1)}\, ,\quad
\epsilon_2=0\, , \quad \epsilon_3=\frac{(\beta +2) (n-2)}{2 (\beta
+2) n-2}\, ,\quad \epsilon_4=\frac{(n-2) (-2 \beta +(\beta +2)
n-1)}{(n-1) ((\beta +2) n-1)}\, .
\end{equation}
By comparing the above indices with the ones in the literature for
the power-law $F(R)$ gravity model (\ref{r2inflation1}), it can be
seen that these coincide for $\beta=0$. In order to calculate the
spectral index, due to the fact that the condition $\epsilon_i\ll 1$
does not necessarily hold true, we need to use
Eq.~(\ref{spectralindex1}) and find the explicit form of $\nu_s$. By
substituting the inflationary indices from
Eq.~(\ref{frgravityconstantroll1r2rnmodel11111}) to Eq.~(\ref{nus}),
the function $\nu_s$ reads,
\begin{equation}
\label{nus1} \nu_s=\frac{3 \beta +2 (\beta +2) n^2-5 (4 \beta +7)
n}{(\beta +2) n^2-(\beta +2) n-1}\, ,
\end{equation}
and by substituting Eq.~(\ref{nus1}) in Eq.~(\ref{spectralindex1}),
the spectral index $n_s$ reads,
\begin{equation}
\label{spectrraindexpowerlaw1} n_s=1-\frac{(n-2) (-3 \beta +(\beta
+2) n-4)}{(\beta +2) n^2-(\beta +2) n-1}\, .
\end{equation}
Accordingly, by substituting the analytic form of $\epsilon_3$ from
Eq.~(\ref{frgravityconstantroll1r2rnmodel11111}) in
Eq.~(\ref{scalartotensor2}), we can obtain the scalar-to-tensor
ratio, which is,
\begin{equation}
\label{scalartotensorpowerlaw} r=\frac{48 (\beta +2)^2 (n-2)^2}{(3
(\beta +2) n-2 (\beta +3))^2} \, .
\end{equation}
It can be crosschecked with the related literature that the
expressions (\ref{spectrraindexpowerlaw1}) and
(\ref{scalartotensorpowerlaw}) coincide with the standard results
for the observational indices corresponding with the slow-roll
power-law $F(R)$ gravity model, when $\beta=0$. Let us investigate
whether the constant-roll power-law $F(R)$ gravity model is viable
or not and we compare the results with the slow-roll power law
power-law $F(R)$ gravity model. With regards to the latter, the
observational indices are identical to the ones appearing in
Eqs.~(\ref{spectrraindexpowerlaw1}) and
(\ref{scalartotensorpowerlaw}) for $\beta=0$ and only for $n=2.3$
the spectral index is in concordance with the Planck data
(\ref{planckdata}), but the scalar-to-tensor ratio is $r=0.284$ so
it is excluded. Let us now see what happens in the constant-roll
case, for which $\beta$ enters the equations, so now we investigate
the parameter space of the model which is governed by the parameters
$(n,\beta)$. In this case, the analysis shows that the compatibility
with observational data can be achieved for a wide range of values
of the parameters. Recall that the spectral index is considered
within the observationally acceptable if it takes values in the
interval $n_s=[0.9595,0.9693]$ and also the scalar-to-tensor ratio
$r$ must satisfy $r<0.07$.
\begin{figure}[h]
\centering
\includegraphics[width=18pc]{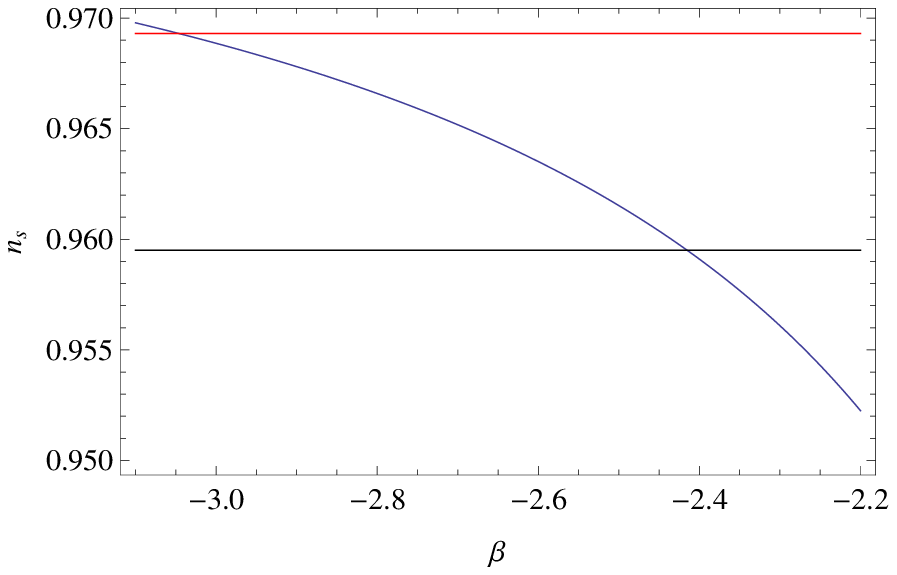}
\includegraphics[width=18pc]{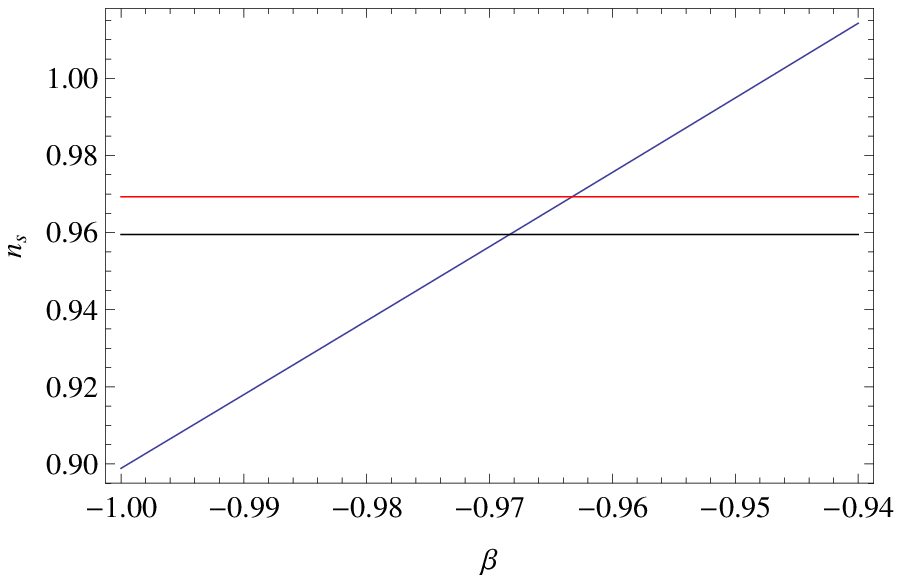}
\caption{The spectral index $n_s=1-\frac{(n-2) (-3 \beta +(\beta +2)
n-4)}{(\beta +2) n^2-(\beta +2) n-1}$ as a function of the parameter
$\beta$, for $n=1.97$ (blue curve, left plot) and for $n=1.1$ (blue
curve, right plot). The red line in both plots corresponds to the
upper bound of the 2015-Planck constraints on the spectral index
$n_s=0.9693$, while the black line in both plots corresponds to the
lower bound of the 2015-Planck data $n_s=0.9595$. }\label{plot3}
\end{figure}
In Fig.~\ref{plot3} we plotted the $\beta$ dependence of the
spectral index for $n=1.97$ (left plot) and for $n=1.1$ (right
plot). As in the previous figures, in both plots, the upper (red)
and lower (black) straight lines correspond to the values
$n_s=0.9693$ and $n_s=0.9595$ respectively. As it can be seen, there
is a large range of values for the parameter $\beta$, for which the
spectral index becomes compatible with observations, however the
analysis of the scalar-to-tensor ratio will reveal some constraints
which should be imposed on the parameter $n$, which recall has to be
positive. So it is useful to use various characteristic examples at
this point. Consider the $\beta=-3$ case, which corresponds to the
ultra-slow-roll scenario of Ref.~\cite{Martin:2012pe}, in which case
for $n=5.24$ the observational indices become,
\begin{equation}
\label{nsr1} n_s=0.966508\, ,\quad r=2.03904\, ,
\end{equation}
and as it can be seen, the scalar-to-tensor value is excluded by
both Planck and BICEP2/Keck-Array data. Consider now the set
$(n,\beta)=(1.983,-2)$, for which the observational indices read,
\begin{equation}
\label{nsr2} n_s=0.966\, ,\quad r=10^{-20}\, ,
\end{equation}
which are both within the Planck and BICEP2/Keck-Array data
constraints, however this scenario is not so appealing since it
predicts an almost zero scalar-to-tensor ratio. Also for
$(n,\beta)=(1.97,-2.75)$ we obtain,
\begin{equation}
\label{nsr3} n_s=0.966\, ,\quad r=0.001\, ,
\end{equation}
which are in good agreement with the Planck and BICEP2/Keck-Array
data constraints. So far all the examples have negative $\beta$, but
there are cases that a positive beta yields optimal results. For
example by choosing $(n,\beta)=(1.89,0.972)$ or equivalently
$(n,\beta)=(2.10149,-0.605095)$, we obtain,
\begin{equation}
\label{nsr4} n_s=0.966\, ,\quad r=0.006\, .
\end{equation}
The above examples show that the large values for the parameter $n$
seem not to be favored, and we discuss this shortly in more detail.
In order to better understand the behavior of the observational
indices as functions of $\beta$ and $n$, we need to provide some
illustrative plots that will clearly demonstrate how the
scalar-to-tensor ratio behaves. In Fig.~\ref{plot4}, we plotted the
scalar-to-tensor ratio $r$ as a function of $\beta$ for $n=1.97$
(left plot), and for $n=3$ (right plot) with the red line in both
cases indicating the BICEP2/Keck-Array constraint $r=0.07$.
Clearly, the right plot reveals an interesting behavior, since the
scalar-to-tensor ratio never becomes compatible with the
BICEP2/Keck-Array constraints. Actually, it is easy to show that for
$n\succeq 2.2$, the scalar-to-tensor ratio never drops below the
bound $r=0.07$ imposed by the BICEP2/Keck-Array collaboration,
regardless of the value of the parameter $n$.

Finally, let us briefly discuss the graceful exit issue, and an
indication that this actually occurs is to find growing perturbation
of the solution (\ref{hubblersquare1}). So consider the following
linear perturbation of the solution (\ref{hubblersquare1}),
\begin{equation}
\label{dhtperturbationlinear} H(t)=\frac{(n-1) ((\beta +2)
n-1)}{(2-n) t}+\Delta H(t)\, ,
\end{equation}
so by substituting (\ref{dhtperturbationlinear}) in
Eq.~(\ref{rsquarebasic1}), we obtain the following differential
equation,
\begin{equation}
\label{afierosimegapempti} \frac{6 (n-1) \left( \left(\beta +2
\right) n-1\right) \left( t \Delta \dot{H}(t)+2 \Delta H(t)\right)
}{t}=0\, ,
\end{equation}
which can be easily solved, with the solution being,
\begin{equation}
\label{dhperturbationsolution} \Delta H(t)=\frac{c_1}{t^2}\, ,
\end{equation}
where $c_1$ is an integration constant. Hence, the solution
(\ref{dhperturbationsolution}) indicates that the linear
perturbations of the solution (\ref{hubblersquare1}) decay as
$t^{-2}$. Interestingly enough, the parameters $n$ and $\beta$ do
not affect the evolution of the perturbations, at least in the
context of the constant-roll approximation, since the differential
equation (\ref{afierosimegapempti}) is obtained by assuming that the
constant-roll condition holds true. We need to note that this result
should be further investigated, since our approach was a linear
perturbation approximation. We hope to address this issue further in
a future work.

\section{Conclusions \label{SecIV}}

The aim of this paper was two-fold, firstly we investigated how
scalar-tensor constant-roll cosmological evolution scenarios can be
realized in the context of vacuum $F(R)$ gravity and secondly we
examined what are the implications of the constant-roll condition on
$F(R)$ gravity, without invoking the scalar-tensor solutions.

In the first approach, after we found the resulting $F(R)$ picture,
we conformally transformed the theory in order to obtain the
Einstein frame scalar-tensor theory. The resulting scalar theory
yields different scalar potential and observational indices in
comparison to the initial scalar theory. We also found the $F(R)$
gravity counterparts of some theories that in the scalar-tensor
frame realize transitions between constant-roll eras. As in the
previous case, the Einstein frame corresponding theory is very
different in comparison to the initial scalar-tensor theory. The
reason behind this difference between the initial scalar theory and
the Einstein frame theory corresponding to the $F(R)$ gravity, is
possibly that the constant-roll condition is quite different in the
Einstein frame in comparison to the $F(R)$ gravity description. This
behavior is also discussed in the literature
\cite{Domenech:2016yxd,Bahamonde:2017kbs,Bahamonde:2016wmz,Brooker:2016oqa},
however for conformal invariant quantities a similarity between the
two frames is expected
\cite{Domenech:2016yxd,Kaiser:1995nv,Faraoni:2007yn}.

In the second approach we investigated the implications of the
constant-roll condition on the inflationary dynamics of a vacuum
$F(R)$ gravity, without invoking the scalar-tensor theory. We
presented the functional form of the inflationary indices in the
constant-roll case, and we compared the results to the standard
slow-roll inflationary indices. After presenting in detail the
formalism of constant-roll $F(R)$ gravity inflation, we applied our
findings in two well known $F(R)$ gravity models, the Starobinsky
and the power-law $F(R)$ gravity model. As we demonstrated, the
Starobinsky model remains compatible with the observational data,
even in the constant-roll case, with the difference in comparison to
the ordinary slow-roll model being that the parameter space is
enlarged, so there is a wide range of parameter values for which the
compatibility with the data can be achieved. In the case of the
power-law $F(R)$ gravity model, the results are more interesting,
since the constant-roll power-law $F(R)$ gravity model of inflation,
can be compatible with observations, in contrast to the slow-roll
one. We performed a thorough analysis in which we studied in detail
the behavior of the spectral index of the power-spectrum of the
primordial curvature perturbations, and of the scalar-to-tensor
ratio, and we examined when the compatibility with the data can be
achieved.

For some future applications we shall mention here that an important
study has to do with the reheating process in the context of $F(R)$
gravity, with the constant-roll inflation governing the inflationary
evolution. This is particularly interesting since it is related to
the graceful exit from inflation issue. It would be interesting to
investigate the implications of a constant-roll era on the reheating
process. Also the issue of non-Gaussianities in the context of
$F(R)$ gravity is also important. The well-known results for
non-Gaussianities that occur if the slow-roll condition is violated,
should in principle hold true in the context of $F(R)$ gravity, too,
but nevertheless, this should be carefully and appropriately
addressed. We hope to study some of these issues in a future work.

Also, after this paper appeared in arXiv, another paper also
appeared in which the constant-roll inflation in $f(R)$ gravity was
studied \cite{Motohashi:2017vdc}. The differences are profound,
since in our case, the constant-roll condition in the Jordan frame
is considered, which is a direct generalization of the scalar-tensor
constant-roll condition. Also in \cite{Motohashi:2017vdc}, the
$f(R)$ gravity is not studied in the Jordan frame but in the
Einstein frame, whereas in our case the second part of this paper is
devoted on the Jordan frame study.

\section*{Acknowledgments}

This work is supported by MINECO (Spain), project FIS2013-44881,
FIS2016-76363-P and by CSIC I-LINK1019 Project (S.D.O) and by
Ministry of Education and Science of Russia (S.D.O and V.K.O), and
(in part) by MEXT KAKENHI Grant-in-Aid for Scientific Research on
Innovative Areas ``Cosmic Acceleration'' (No. 15H05890) (S.N.).


\begin{thebibliography}{99}

\bibitem{Linde:2007fr}
A.~D.~Linde,
Lect.\ Notes Phys.\  {\bf 738} (2008) 1
doi:10.1007/978-3-540-74353-8\_1 [arXiv:0705.0164 [hep-th]].

\bibitem{Gorbunov:2011zzc}
D.~S.~Gorbunov and V.~A.~Rubakov,``Introduction to the theory of the
early universe: Cosmological perturbations and inflationary
theory,'' Hackensack, USA: World Scientific (2011) 489 p

\bibitem{Lyth:1998xn}
D.~H.~Lyth and A.~Riotto,
Phys.\ Rept.\  {\bf 314} (1999) 1 doi:10.1016/S0370-1573(98)00128-8
[hep-ph/9807278].



\bibitem{Ade:2015lrj}
P.~A.~R.~Ade {\it et al.} [Planck Collaboration],
Astron.\ Astrophys.\  {\bf 594} (2016) A20
doi:10.1051/0004-6361/201525898 [arXiv:1502.02114 [astro-ph.CO]].



\bibitem{Martin:2016ckm}
  J.~Martin,
  Class.\ Quant.\ Grav.\  {\bf 33} (2016) no.3,  034001.
  doi:10.1088/0264-9381/33/3/034001






\bibitem{Starobinsky:1980te}
A.~A.~Starobinsky,
Phys.\ Lett.\  {\bf 91B} (1980) 99. doi:10.1016/0370-2693(80)90670-X

\bibitem{Barrow:1988xh}
J.~D.~Barrow and S.~Cotsakis,
Phys.\ Lett.\ B {\bf 214} (1988) 515.
doi:10.1016/0370-2693(88)90110-4

\bibitem{Bezrukov:2007ep}
F.~L.~Bezrukov and M.~Shaposhnikov,
Phys.\ Lett.\ B {\bf 659} (2008) 703
doi:10.1016/j.physletb.2007.11.072 [arXiv:0710.3755 [hep-th]].

\bibitem{Kallosh:2013hoa}
R.~Kallosh and A.~Linde,
JCAP {\bf 1307} (2013) 002 doi:10.1088/1475-7516/2013/07/002
[arXiv:1306.5220 [hep-th]].

\bibitem{Ferrara:2013rsa}
S.~Ferrara, R.~Kallosh, A.~Linde and M.~Porrati,
Phys.\ Rev.\ D {\bf 88} (2013) no.8,  085038
doi:10.1103/PhysRevD.88.085038 [arXiv:1307.7696 [hep-th]].

\bibitem{Kallosh:2013yoa}
R.~Kallosh, A.~Linde and D.~Roest,
JHEP {\bf 1311} (2013) 198 doi:10.1007/JHEP11(2013)198
[arXiv:1311.0472 [hep-th]].

\bibitem{Odintsov:2016vzz}
S.~D.~Odintsov and V.~K.~Oikonomou,
Phys.\ Rev.\ D {\bf 94} (2016) no.12,  124026
doi:10.1103/PhysRevD.94.124026 [arXiv:1612.01126 [gr-qc]].

\bibitem{Odintsov:2016jwr}
S.~D.~Odintsov and V.~K.~Oikonomou,
arXiv:1611.00738 [gr-qc].

\bibitem{Chen:2010xka}
X.~Chen,
Adv.\ Astron.\  {\bf 2010} (2010) 638979 doi:10.1155/2010/638979
[arXiv:1002.1416 [astro-ph.CO]].

\bibitem{Inoue:2001zt}
S.~Inoue and J.~Yokoyama,
Phys.\ Lett.\ B {\bf 524} (2002) 15
doi:10.1016/S0370-2693(01)01369-7 [hep-ph/0104083].

\bibitem{Tsamis:2003px}
N.~C.~Tsamis and R.~P.~Woodard,
Phys.\ Rev.\ D {\bf 69} (2004) 084005 doi:10.1103/PhysRevD.69.084005
[astro-ph/0307463].

\bibitem{Kinney:2005vj}
W.~H.~Kinney,
Phys.\ Rev.\ D {\bf 72} (2005) 023515 doi:10.1103/PhysRevD.72.023515
[gr-qc/0503017].

\bibitem{Tzirakis:2007bf}
K.~Tzirakis and W.~H.~Kinney,
Phys.\ Rev.\ D {\bf 75} (2007) 123510 doi:10.1103/PhysRevD.75.123510
[astro-ph/0701432].

\bibitem{Namjoo:2012aa}
M.~H.~Namjoo, H.~Firouzjahi and M.~Sasaki,
Europhys.\ Lett.\  {\bf 101} (2013) 39001
doi:10.1209/0295-5075/101/39001 [arXiv:1210.3692 [astro-ph.CO]].

\bibitem{Martin:2012pe}
J.~Martin, H.~Motohashi and T.~Suyama,
Phys.\ Rev.\ D {\bf 87} (2013) no.2,  023514
doi:10.1103/PhysRevD.87.023514 [arXiv:1211.0083 [astro-ph.CO]].

\bibitem{Motohashi:2014ppa}
H.~Motohashi, A.~A.~Starobinsky and J.~Yokoyama,
JCAP {\bf 1509} (2015) no.09,  018 doi:10.1088/1475-7516/2015/09/018
[arXiv:1411.5021 [astro-ph.CO]].

\bibitem{Cai:2016ngx}
Y.~F.~Cai, J.~O.~Gong, D.~G.~Wang and Z.~Wang,
JCAP {\bf 1610} (2016) no.10,  017 doi:10.1088/1475-7516/2016/10/017
[arXiv:1607.07872 [astro-ph.CO]].

\bibitem{Motohashi:2017aob}
H.~Motohashi and A.~A.~Starobinsky,
arXiv:1702.05847 [astro-ph.CO].

\bibitem{Hirano:2016gmv}
S.~Hirano, T.~Kobayashi and S.~Yokoyama,
Phys.\ Rev.\ D {\bf 94} (2016) no.10,  103515
doi:10.1103/PhysRevD.94.103515 [arXiv:1604.00141 [astro-ph.CO]].

\bibitem{Anguelova:2015dgt}
L.~Anguelova,
Nucl.\ Phys.\ B {\bf 911} (2016) 480
doi:10.1016/j.nuclphysb.2016.08.020 [arXiv:1512.08556 [hep-th]].

\bibitem{Cook:2015hma}
J.~L.~Cook and L.~M.~Krauss,
JCAP {\bf 1603} (2016) no.03,  028 doi:10.1088/1475-7516/2016/03/028
[arXiv:1508.03647 [astro-ph.CO]].

\bibitem{Kumar:2015mfa}
K.~S.~Kumar, J.~Marto, P.~Vargas Moniz and S.~Das,
JCAP {\bf 1604} (2016) no.04,  005 doi:10.1088/1475-7516/2016/04/005
[arXiv:1506.05366 [gr-qc]].

\bibitem{Odintsov:2017yud}
S.~D.~Odintsov and V.~K.~Oikonomou,
arXiv:1703.02853 [gr-qc].

\bibitem{Odintsov:2017qpp}
S.~D.~Odintsov and V.~K.~Oikonomou,
arXiv:1704.02931 [gr-qc].

\bibitem{Lin:2015fqa}
J.~Lin, Q.~Gao and Y.~Gong,
Mon.\ Not.\ Roy.\ Astron.\ Soc.\  {\bf 459} (2016) no.4,  4029
doi:10.1093/mnras/stw915 [arXiv:1508.07145 [gr-qc]].

\bibitem{Gao:2017uja}
Q.~Gao and Y.~Gong,
arXiv:1703.02220 [gr-qc].

\bibitem{Nojiri:2006ri}
S.~Nojiri and S.~D.~Odintsov,
eConf C {\bf 0602061} (2006) 06 [Int.\ J.\ Geom.\ Meth.\ Mod.\
Phys.\ {\bf 4} (2007) 115] doi:10.1142/S0219887807001928
[hep-th/0601213].

\bibitem{Capozziello:2011et}
S.~Capozziello and M.~De Laurentis,
Phys.\ Rept.\ {\bf 509} (2011) 167 doi:10.1016/j.physrep.2011.09.003
[arXiv:1108.6266 [gr-qc]].

\bibitem{Capozziello:2010zz}
V.~Faraoni and S.~Capozziello,
Fundam.\ Theor.\ Phys.\ {\bf 170} (2010).
doi:10.1007/978-94-007-0165-6

\bibitem{Capozziello:2009nq}
S.~Capozziello, M.~De Laurentis and V.~Faraoni,
Open Astron.\ J.\ {\bf 3} (2010) 49 doi:10.2174/1874381101003010049,
10.2174/1874381101003020049 [arXiv:0909.4672 [gr-qc]].

\bibitem{Nojiri:2010wj}
S.~Nojiri and S.~D.~Odintsov,
Phys.\ Rept.\ {\bf 505} (2011) 59 doi:10.1016/j.physrep.2011.04.001
[arXiv:1011.0544 [gr-qc]].

\bibitem{Clifton:2011jh}
T.~Clifton, P.~G.~Ferreira, A.~Padilla and C.~Skordis,
Phys.\ Rept.\ {\bf 513} (2012) 1
[arXiv:1106.2476 [astro-ph.CO]].

\bibitem{Nojiri:2006gh}
S.~Nojiri and S.~D.~Odintsov,
Phys.\ Rev.\ D {\bf 74} (2006) 086005 doi:10.1103/PhysRevD.74.086005
[hep-th/0608008].

\bibitem{Domenech:2016yxd}
G.~Domenech and M.~Sasaki,
Int.\ J.\ Mod.\ Phys.\ D {\bf 25} (2016) no.13, 1645006
doi:10.1142/S0218271816450061 [arXiv:1602.06332 [gr-qc]].

\bibitem{Bahamonde:2017kbs}
S.~Bahamonde, S.~D.~Odintsov, V.~K.~Oikonomou and P.~V.~Tretyakov,
Phys.\ Lett.\ B {\bf 766} (2017) 225
doi:10.1016/j.physletb.2017.01.012 [arXiv:1701.02381 [gr-qc]].

\bibitem{Bahamonde:2016wmz}
S.~Bahamonde, S.~D.~Odintsov, V.~K.~Oikonomou and M.~Wright,
Annals Phys.\  {\bf 373} (2016) 96 doi:10.1016/j.aop.2016.06.020
[arXiv:1603.05113 [gr-qc]].


\bibitem{Brooker:2016oqa}
  D.~J.~Brooker, S.~D.~Odintsov and R.~P.~Woodard,
  Nucl.\ Phys.\ B {\bf 911} (2016) 318
  doi:10.1016/j.nuclphysb.2016.08.010
  [arXiv:1606.05879 [gr-qc]].








\bibitem{Array:2015xqh}
P.~A.~R.~Ade {\it et al.} [BICEP2 and Keck Array Collaborations],
Phys.\ Rev.\ Lett.\ {\bf 116} (2016) 031302
doi:10.1103/PhysRevLett.116.031302 [arXiv:1510.09217 [astro-ph.CO]].

\bibitem{Noh:2001ia}
H.~Noh and J.~c.~Hwang,
Phys.\ Lett.\ B {\bf 515} (2001) 231
[astro-ph/0107069].

\bibitem{Hwang:2001qk}
J.~c.~Hwang and H.~r.~Noh,
Phys.\ Rev.\ D {\bf 65} (2002) 023512 doi:10.1103/PhysRevD.65.023512
[astro-ph/0102005].

\bibitem{Hwang:2001pu}
J.~c.~Hwang and H.~Noh,
Phys.\ Lett.\ B {\bf 506} (2001) 13
doi:10.1016/S0370-2693(01)00404-X [astro-ph/0102423].

\bibitem{Nojiri:2016vhu}
S.~Nojiri, S.~D.~Odintsov and V.~K.~Oikonomou,
Phys.\ Rev.\ D {\bf 94} (2016) no.10,  104050
doi:10.1103/PhysRevD.94.104050 [arXiv:1608.07806 [gr-qc]].

\bibitem{Odintsov:2016plw}
S.~D.~Odintsov and V.~K.~Oikonomou,
Class.\ Quant.\ Grav.\  {\bf 33} (2016) no.12,  125029
doi:10.1088/0264-9381/33/12/125029 [arXiv:1602.03309 [gr-qc]].

\bibitem{Odintsov:2015gba}
S.~D.~Odintsov and V.~K.~Oikonomou,
Phys.\ Rev.\ D {\bf 92} (2015) no.12,  124024
doi:10.1103/PhysRevD.92.124024 [arXiv:1510.04333 [gr-qc]].

\bibitem{Bamba:2014jia}
K.~Bamba, R.~Myrzakulov, S.~D.~Odintsov and L.~Sebastiani,
Phys.\ Rev.\ D {\bf 90} (2014) no.4, 043505
doi:10.1103/PhysRevD.90.043505 [arXiv:1403.6649 [hep-th]].

\bibitem{Kaiser:1995nv}
D.~I.~Kaiser,
[astro-ph/9507048].

\bibitem{Faraoni:2007yn}
V.~Faraoni,
Phys.\ Rev.\ D {\bf 75} (2007) 067302 doi:10.1103/PhysRevD.75.067302
[gr-qc/0703044 [GR-QC]].




\bibitem{Motohashi:2017vdc}
  H.~Motohashi and A.~A.~Starobinsky,
  Eur.\ Phys.\ J.\ C {\bf 77} (2017) no.8,  538
  doi:10.1140/epjc/s10052-017-5109-x
  [arXiv:1704.08188 [astro-ph.CO]].




\end{thebibliography}
\end{document}